

\def\nofirstpagenoten{\nopagenumbers\footline={\ifnum\pageno>1\tenrm
\hss\folio\hss\fi}}
\def\nofirstpagenotwelve{\nopagenumbers\footline={\ifnum\pageno>1\twelverm
\hss\folio\hss\fi}}
\def\leaderfill{\leaders\hbox to 1em{\hss.\hss}\hfill}

\def\frac#1/#2{\leavevmode\kern.1em
\raise.5ex\hbox{\the\scriptfont0 #1}\kern-.1em/\kern-.15em
\lower.25ex\hbox{\the\scriptfont0 #2}}
\def\sfrac#1/#2{\leavevmode\kern.1em
\raise.5ex\hbox{\the\scriptscriptfont0 #1}\kern-.1em/\kern-.15em
\lower.25ex\hbox{\the\scriptscriptfont0 #2}}


\parindent=20pt
\def\narrow{\advance\leftskip by 40pt \advance\rightskip by 40pt}

\def\nonarrower{\advance\leftskip by -40pt\advance\rightskip by -40pt}

\def\boxit#1{\vbox{\hrule\hbox{\vrule\kern3pt
        \vbox{\kern3pt#1\kern3pt}\kern3pt\vrule}\hrule}}

\def\gtorder{\mathrel{\raise.3ex\hbox{$>$}\mkern-14mu
             \lower0.6ex\hbox{$\sim$}}}
\def\ltorder{\mathrel{\raise.3ex\hbox{$<$}|mkern-14mu
             \lower0.6ex\hbox{\sim$}}}
\def\dalemb#1#2{{\vbox{\hrule height .#2pt
        \hbox{\vrule width.#2pt height#1pt \kern#1pt
                \vrule width.#2pt}
        \hrule height.#2pt}}}

\font\fourteentt=cmtt10 scaled \magstep2
\font\fourteenbf=cmbx12 scaled \magstep1
\font\fourteenrm=cmr12 scaled \magstep1
\font\fourteeni=cmmi12 scaled \magstep1
\font\fourteenss=cmss12 scaled \magstep1
\font\fourteensy=cmsy10 scaled \magstep2
\font\fourteensl=cmsl12 scaled \magstep1
\font\fourteenex=cmex10 scaled \magstep2
\font\fourteenit=cmti12 scaled \magstep1
\font\twelvett=cmtt10 scaled \magstep1 \font\twelvebf=cmbx12
\font\twelverm=cmr12 \font\twelvei=cmmi12
\font\twelvess=cmss12 \font\twelvesy=cmsy10 scaled \magstep1
\font\twelvesl=cmsl12 \font\twelveex=cmex10 scaled \magstep1
\font\twelveit=cmti12
\font\tenss=cmss10
 
 \font\ninebf=cmbx7 scaled \magstep1
\font\ninerm=cmr7 scaled \magstep1 \font\ninei=cmmi7 scaled \magstep1
\font\ninesy=cmsy7 scaled \magstep1 
\font\eightrm=cmr7 scaled 1140 
 
\font\sevenbf=cmbx7 \font\sevenrm=cmr7 \font\seveni=cmmi7
\font\sevensy=cmsy7 

\catcode`@=11
\newskip\ttglue
\newfam\ssfam

\def\fourteenpoint{\def\rm{\fam0\fourteenrm}
\textfont0=\fourteenrm \scriptfont0=\tenrm \scriptscriptfont0=\sevenrm
\textfont1=\fourteeni \scriptfont1=\teni \scriptscriptfont1=\seveni
\textfont2=\fourteensy \scriptfont2=\tensy \scriptscriptfont2=\sevensy
\textfont3=\fourteenex \scriptfont3=\fourteenex \scriptscriptfont3=\fourteenex
\def\it{\fam\itfam\fourteenit} \textfont\itfam=\fourteenit
\def\sl{\fam\slfam\fourteensl} \textfont\slfam=\fourteensl
\def\bf{\fam\bffam\fourteenbf} \textfont\bffam=\fourteenbf
\scriptfont\bffam=\tenbf \scriptscriptfont\bffam=\sevenbf
\def\tt{\fam\ttfam\fourteentt} \textfont\ttfam=\fourteentt
\def\ss{\fam\ssfam\fourteenss} \textfont\ssfam=\fourteenss
\tt \ttglue=.5em plus .25em minus .15em
\normalbaselineskip=16pt
\abovedisplayskip=16pt plus 4pt minus 12pt
\belowdisplayskip=16pt plus 4pt minus 12pt
\abovedisplayshortskip=0pt plus 4pt
\belowdisplayshortskip=9pt plus 4pt minus 6pt
\parskip=5pt plus 1.5pt
\setbox\strutbox=\hbox{\vrule height12pt depth5pt width0pt}
\let\sc=\tenrm
\let\big=\fourteenbig \normalbaselines\rm}
\def\fourteenbig#1{{\hbox{$\left#1\vbox to12pt{}\right.\n@space$}}}

\def\twelvepoint{\def\rm{\fam0\twelverm}
\textfont0=\twelverm \scriptfont0=\ninerm \scriptscriptfont0=\sevenrm
\textfont1=\twelvei \scriptfont1=\ninei \scriptscriptfont1=\seveni
\textfont2=\twelvesy \scriptfont2=\ninesy \scriptscriptfont2=\sevensy
\textfont3=\twelveex \scriptfont3=\twelveex \scriptscriptfont3=\twelveex
\def\it{\fam\itfam\twelveit} \textfont\itfam=\twelveit
\def\sl{\fam\slfam\twelvesl} \textfont\slfam=\twelvesl
\def\bf{\fam\bffam\twelvebf} \textfont\bffam=\twelvebf
\scriptfont\bffam=\ninebf \scriptscriptfont\bffam=\sevenbf
\def\tt{\fam\ttfam\twelvett} \textfont\ttfam=\twelvett
\def\ss{\fam\ssfam\twelvess} \textfont\ssfam=\twelvess
\tt \ttglue=.5em plus .25em minus .15em
\normalbaselineskip=14pt
\abovedisplayskip=14pt plus 3pt minus 10pt
\belowdisplayskip=14pt plus 3pt minus 10pt
\abovedisplayshortskip=0pt plus 3pt
\belowdisplayshortskip=8pt plus 3pt minus 5pt
\parskip=3pt plus 1.5pt
\setbox\strutbox=\hbox{\vrule height10pt depth4pt width0pt}
\let\sc=\ninerm
\let\big=\twelvebig \normalbaselines\rm}
\def\twelvebig#1{{\hbox{$\left#1\vbox to10pt{}\right.\n@space$}}}

\def\tenpoint{\def\rm{\fam0\tenrm}
\textfont0=\tenrm \scriptfont0=\sevenrm \scriptscriptfont0=\fiverm
\textfont1=\teni \scriptfont1=\seveni \scriptscriptfont1=\fivei
\textfont2=\tensy \scriptfont2=\sevensy \scriptscriptfont2=\fivesy
\textfont3=\tenex \scriptfont3=\tenex \scriptscriptfont3=\tenex
\def\it{\fam\itfam\tenit} \textfont\itfam=\tenit
\def\sl{\fam\slfam\tensl} \textfont\slfam=\tensl
\def\bf{\fam\bffam\tenbf} \textfont\bffam=\tenbf
\scriptfont\bffam=\sevenbf \scriptscriptfont\bffam=\fivebf
\def\tt{\fam\ttfam\tentt} \textfont\ttfam=\tentt
\def\ss{\fam\ssfam\tenss} \textfont\ssfam=\tenss
\tt \ttglue=.5em plus .25em minus .15em
\normalbaselineskip=12pt
\abovedisplayskip=12pt plus 3pt minus 9pt
\belowdisplayskip=12pt plus 3pt minus 9pt
\abovedisplayshortskip=0pt plus 3pt
\belowdisplayshortskip=7pt plus 3pt minus 4pt
\parskip=0.0pt plus 1.0pt
\setbox\strutbox=\hbox{\vrule height8.5pt depth3.5pt width0pt}
\let\sc=\eightrm
\let\big=\tenbig \normalbaselines\rm}
\def\tenbig#1{{\hbox{$\left#1\vbox to8.5pt{}\right.\n@space$}}}
\let\rawfootnote=\footnote \def\footnote#1#2{{\rm\parskip=0pt\rawfootnote{#1}
{#2\hfill\vrule height 0pt depth 6pt width 0pt}}}

 \magnification 1200 \baselineskip=18pt \def \h{{1\over 2}} \def\a{A} \def\b{K}
\def\p{\partial} \overfullrule=0pt \def\dg{$^\dagger$} 
\line{\hfil CTP-TAMU-101/91} \vskip 1.0truein

\centerline{\bf THE COUPLING OF YANG-MILLS TO EXTENDED OBJECTS} \vskip
.90truein

\centerline{J. A. Dixon, M. J. Duff{\footnote\dg {Work supported in part by NSF
grant PHY-9106593}} and E. Sezgin \dg} \bigskip

\centerline{Center for Theoretical Physics} \centerline{Physics Department}
\centerline{Texas A\&M University} \centerline{College Station, Texas 77843}
\vskip .4truein \centerline{\bf  December 1991} \vskip .5truein

\centerline{\bf ABSTRACT} \medskip

 The coupling of Yang-Mills fields to the heterotic string in bosonic
formulation is generalized to extended objects of higher dimension (p-branes).
For odd $p$, the Bianchi identities obeyed by the field strengths of the
$(p+1)$-forms receive Chern-Simons corrections which,  in the case  of the
5--brane, are consistent with an earlier conjecture based on string/5-brane
duality. \vfill\eject

\noindent {\bf I. Introduction} \medskip

	Although the effort to generalize the physics of superstrings to
higher-dimensional objects, super-p-branes, has been an active area of research
since 1986, it is only recently that attention has turned to incorporating
internal symmetries. The heterotic string [1] provides the paradigm for such
Yang-Mills couplings, and here the problem is well understood. We have the
luxury of employing either a fermionic formulation in which chiral fermions on
the 2-dimensional worldsheet carry the internal quantum numbers or a bosonic
formulation where the basic variables can be either free bosons or else the
coordinates on a simply-laced group manifold.  To date, no analogous action has
been found for p-branes even though the existence of a ``heterotic fivebrane''
was conjectured in 1987 [2].  However, now at least we have an existence proof:
the heterotic fivebrane emerges as a soliton solution of the heterotic string
[3]. A study of the zero-modes of this soliton suggests that the group manifold
approach might be a good starting point for constructing the action.  Here one
must distinguish between the covariant Green-Schwarz action and the gauge-fixed
action that decribes only physical degrees of freedom. In the former case, the
problem is to generalize the D=10 spacetime supersymmetric and
$\kappa$-invariant fivebrane action of [4] to include the internal degrees of
freedom which correspond presumably to the group manifold of $SO(32)$ or $ E_8
\times E_8$.  In the latter case it is to find an action supersymmetric on the
d=6 worldvolume, which would involve a non-linear $ \sigma$ model of a
quaternionic Kahler manifold.

	In this paper, we make a first step toward the construction of the
heterotic fivebrane by adopting the group manifold approach to coupling
Yang-Mills fields to bosonic extended objects. For generality, we consider a
d-dimensional ($d=p+1$) worldvolume and a D-dimensional spacetime. Let us begin
by reviewing the bosonic sector of the heterotic string. \bigskip \noindent{\bf
2. Coupling Yang-Mills Field to the String} \bigskip The bosonic sector of the
heterotic string may be described by the action $S_2=S_2^K+S_2^W$, where [5] $$
 \eqalignno{  S_2^K&=\int d^2 \xi\bigg\{ -\h \sqrt{-\gamma}
\gamma^{ij}\bigg(\p_i X^\mu\p_j X^\nu g_{\mu\nu}(X)+\p_i y^m \p_j y^n
g_{mn}(y)\bigg)\bigg\}  &(2.1) \cr  S_2^W&=  \int d^2 \xi\bigg\{
-\h\epsilon^{ij}\bigg(-\p_i X^\mu\p_j X^\nu B_{\mu\nu}(X)  +\p_i y^m \p_j y^n
b_{mn}(y)\bigg) \bigg\} &(2.2) \cr} $$  where $\xi^i\ (i=0,1)$ are the
worldsheet coordinates, $x^\mu(\xi)\ (\mu=0,...,9)$ are the spacetime
coordinates and $\gamma_{ij}(\xi)$ the worldsheet metric\footnote{$^\dagger$}{
Here, and in the rest of the paper, we set  the  dimensionful parameters as
well as the possibly quantized coupling constants equal to  one.}. The first
terms in $S_2^K$ and $S_2^W$ are just the usual Green-Schwarz couplings to the
background spacetime metric $g_{\mu\nu}(X)$ and rank-2 antisymmetric tensor
$B_{\mu\nu}(X)$. The second term in $S_2^K$ describes a nonlinear
$\sigma$-model on the compact semi-simple Lie group manifold G, where
$y^m(\xi)\ (m=1,..., {\rm dim}\  G)$ are the coordinates on G and $g_{mn}(y)$
is the bi-invariant metric. Introducing the left-invariant Killing vectors
$K^a_m(y)$, we have \footnote{${\dagger\dagger}$}{In our conventions, the
generators of the group obey the algebra $[T_a, T_b]=f_{ab}^c T_c$.   The
raising and lowering of indices will be done with the invariant tensor $d_{ab}$
defined by \break\hfill tr $T_a T_b  = d_{ab}$.} $$ \eqalign{
g_{mn}&=K^a_m K^a_n  \cr             \p_m K^a_n-\p_n K^a_m &= -f^a_{bc} K^b_m
K^c_n ,  \cr}\eqno(2.3) $$  The second term in $S_2^W$ is the WZW term,
involving the rank-2 tensor $b_{mn}(y)$, for which $$  h_{mnp} \equiv  3
\p_{[m}b_{np]} -  f_{abc}K^a_m K^b_n K^c_p = 0   \eqno(2.4)    $$    Strictly
speaking, for the string to be heterotic, we require that the bosons $y^m(\xi)$
be chiral on the d=2 worldsheet. This is also required for $\kappa$-symmetry.
Since in this paper we are primarily concerned with the bosonic sector of
$(d-1)$-branes with $d\ge 2$, we shall omit this constraint. The action is
invariant under rigid $G_L\times G_R $ transformations. For $G_L$ they are $$
 \delta y^m=K^a_m(y) \lambda^a   \eqno(2.5)  $$ In gauging $G_L$, however, by
allowing $ \lambda^a= \lambda^a(X)$, there is a subtlety. In the kinetic term
$S_2^K$ it is sufficient to introduce the covariantly transforming currents $$
J^a_i=\p_i X^\mu A^a_\mu-\p_i y^m K^a_m   \eqno(2.6) $$ where $A_\mu^a (X)$ are
the Yang-Mills gauge-fields transforming as $$   \delta A_\mu^a=\p_\mu
\lambda^a+f^a_{bc}A_\mu^b  \lambda^c   \eqno(2.7)   $$   Thus  the gauge
invariant extension of (2.1) is  $$   S^K_2 = \int d^2\xi  \bigg\{-\h
\sqrt{-\gamma} \gamma^{ij}\bigg(\p_i X^\mu \p_j X^\nu g_{\mu\nu}+J^a_i J^a_j
\bigg)\bigg\}  \eqno(2.8) $$ In the WZW term, however, we have three terms [6]
$$   S^W_2= \int d^2\xi\bigg\{-\h\epsilon^{ij}\bigg( -\p_i X^\mu \p _j X^\nu
B_{\mu\nu}-   2\p_i X^\mu A_\mu^a\p_j y^m K^a_m + \p_i y^m \p_j y^n
b_{mn}\bigg)\bigg\} \eqno(2.9)   $$  and the invariance can be achieved only by
assigning a non-trivial transformation rule to the rank-2 tensor $B_{\mu\nu}$,
namely $$    \delta B_{\mu\nu}=- A_\mu^a\p_\nu  \lambda^a+A_\nu^a\p_\mu
\lambda^a  \eqno(2.10) $$

In order to  generalize this construction to d-dimensional extended objects in
D spacetime dimensions, it is useful to adopt a condensed notation to rewrite
the string action in terms of  building blocks which  may  readily admit higher
dimensional generalizations.   Introduce the Lie-algebra valued 1-forms $$
\eqalignno{      \a &= A_\mu^a (X) T^a \p_i X^\mu d\xi^i  &(2.11)\cr
\b &= K^a_m(y) T ^a\p_i y^m d\xi^i  &(2.12)\cr} $$ where $T^a$ are the
generators of $G$  in the fundamental representation.  As a consequence of
(2.3), we have the Maurer-Cartan equation $$ d\b +\b^2=0, \eqno(2.13) $$ where
$d$ is the exterior derivative $d=d \xi^i   \p_i = d \xi^i {\p\over \p\xi^i}$.
Furthermore,  using (2.5) and (2.7), the gauge transformations  of $\a$ and
$\b$ can be expressed as follows $$ \eqalignno{    \delta\a &= d \lambda +[\a,
\lambda]  &(2.14)\cr     \delta\b &= d \lambda +[\b, \lambda]  &(2.15)\cr} $$
Note that the same parameter $ \lambda$ occurs in both of the transformation
rules.  As a consequence of this  the combination $\a-\b$ transforms
covariantly  $$  \delta(\a-\b)=[\a-\b,  \lambda]  \eqno(2.16)  $$  In fact,  $$
          J^a_i T^a d\xi^i=\a-\b \equiv J \eqno(2.17) $$  which makes manifest
the covariant transformation character of $J^a_i$, and hence the gauge
invariance of the kinetic action (2.8).

In order to write the WZW action in a compact form as well,  let us also define
the $d$-forms $$    \eqalign{        B_{d} &= {1\over d!} B_{\mu_1...\mu_d
}\p_{i_1} X^{\mu_1}\cdots \p_{i_d} X^{\mu_d} d\xi^{i_1}\cdots        d
\xi^{i_d} \cr         b_d &={1\over d!}  b_{m_1...m_d}\p_{i_1} y^{m_1}\cdots
\p_{i_d} y^{m_d}  d\xi^{i_1}\cdots         d \xi^{i_d} \cr}   \eqno(2.18) $$
Then the   WZW action (2.9) may be written $$    S_2^W = \int \big\{ B_2 +{\rm
tr} ( \a\b) -b_2\big\}  \eqno(2.19) $$ It is useful to introduce the notation
$$     C_2 \equiv {\rm  tr}( \a\b)  \eqno(2.20) $$ Here, and in the rest of the
paper, $tr$  refers to trace in the {\it fundamental} representation.  The
gauge invariance of $S^W_2$ can now be understood as follows. First, consider
the gauge invariant polynomial $I_4(F)$, where $F= dA +A^2$. We then note the
usual descent equations $$ I_4(F) = d I^0_3(F,A) $$ $$ \delta I_3(F,A) = d
I^1_2(F,A,\lambda) \eqno(2.21) $$ The subscripts on $I$ denote form degree and
the superscripts count the number of gauge parameters $\lambda$. Next, in
condensed notation (2.4) reads $$ h_3 \equiv  d b_2 + I^0_3(\b) = 0 \eqno(2.22)
$$ and hence    (up to a total derivative term) $$   \eqalign{
    \delta b_2  &=-{\rm tr} (\b d \lambda)  \cr
           &\equiv -I_2^1(\b, \lambda)   \cr} \eqno (2.23) $$  Then, from
(2.10) and (2.11) we have $$ \eqalign{                     \delta B_2=- {\rm
tr} (\a d \lambda ) \cr                                            \equiv
-I_2^1(\a)  \cr}\eqno(2.24) $$   The total derivative term which we have
dropped in (2.23)  corresponds to a tensor gauge transformation of $b_2$. The
action  is,  of course, invariant under these tensor gauge transformations as
well as  similar tensor gauge transformation of $B_2$.  In the rest of this
paper,  we shall focus  on the Yang-Mills gauge transformations.    Finally the
gauge transformation of  $C_2$ is easily found to be   $$             \delta
C_2 =I_2^1(\a, \lambda)-I_2^1 (\b, \lambda)   \eqno(2.25) $$ The manner in
which the WZW action (2.19) is gauge invariant is now transparent,  given the
transformation rules (2.23), (2.24) and (2.25). \bigskip \noindent{\bf 3.
Coupling of Yang-Mills Field to the  Higher Dimensional Extended Objects}
\bigskip The background fields in this case  are the metric $g_{\mu\nu}\
(\mu=0,... D-1)$ and a rank-d antisymmetric tensor $B_{\mu_1...\mu_d} (X)$.
The generalization of the kinetic term (2.8) is obvious, namely $$ S^K_d =\int
d^d \xi  \bigg\{-\h \sqrt{-\gamma} \gamma^{ij}\bigg(\p_i X^\mu \p_j X^\nu
g_{\mu\nu} +J^a_i J^a_j \bigg)  +{1\over 2}(d-2) \sqrt{-\gamma} \bigg\}
\eqno(3.1) $$ where $\xi^i\  (i=0,...d-1)$ are the worldvolume coordinates. In
order to generalize the results of the last section to higher dimensions,
consider a gauge invariant polynomial $I_{2n+2}(F)$ in  $(2n+2)$-dimensions.
Since $I_{2n+2}(F)$ is closed and gauge invariant, we have the descent
equations $$     \eqalign {                    I_{2n+2}(F) &= d
I_{2n+1}^0(F,\a)  \cr                       \delta I_{2n+1}^0(F,\a) &= d
I_{2n}^1( F,\a, \lambda) \cr}, \eqno(3.2) $$ Note that   $I_{2n+2}(F)$  is an
even form,  and hence these descent equations are relevant  for $p$-branes with
$p$ odd,  since $d=p+1=2n$.   We shall come back to the case of   $p$-branes
with $p$ even.  Note also that, since the curvature of $\b$ is vanishing,
$I_{2n+1}^0$ is an algebraic polynomial in $\b$, and $d I_{2n+1}^0(\b)=0$.   In
analogy with (2.19) we propose the following  WZW action $$    S_{2n}^W =\int
\big\{ B_{2n} +C_{2n}(\a,\b) -b_{2n}  \big\}\equiv \int {\cal B}_{2n}
\eqno(3.3) $$ where  $C_{2n} (\a,\b)$, the  analog of $C_2$ given in (2.20), is
still to be determined.  $b_{2n}$ is  again   chosen so that $$ h_{2n+1}
\equiv db_{2n}+ I^0_{2n+1}(K)=0  \eqno(3.4) $$  The  non-zero Chern-Simons
forms $I^0_{2n+1}(K)$ are in one to one correspondence with the non-zero
totally antisymmetric  group invariant tensors of the group $G$.  These are in
turn generated by products of the primitive antisymmetric tensors of the group,
which are (nearly) all of the form $ {\rm tr} T^{[a_1} \cdots T^{a_{(2m +1)}]
}$. Tables of the cohomology of the Lie algebras tell us which of these tensors
are non-zero [7].    For example, we can construct an  $SU(3)$ invariant
3-brane using $I^0_{5} = a_5 {\rm tr} K^5 $, and we can construct an  $SO(2N)$
invariant 5-brane using $I^0_{7} = a_7 {\rm tr} K^7$. Here $a_{2n+1}$ are
calculable constants. In some cases the WZW term $b_{2n}$ does not exist, e.g.
$G=SO(2N)$ (except for $N=3$) for the 3-brane and $G=E_8\times E_8$ for the
5-brane.  In the case of string, we know that global considerations play a role
and yield a quantization condition on the coefficient of the WZW term [5]. We
intend to return to such global questions for $p$-branes elsewhere.

  The $(2n+1)$ forms  $db_{2n}$ and $I_{2n +1}$ are    of course defined on a
$(2n+1) $ dimensional space whose boundary is the $2n$ dimensional worldvolume.
A derivation similar to that of (2.23) yields the following transformation rule
for   $b_{2n}$ $$   \delta b_{2n} = -I_{2n}^1(\b, \lambda)   \eqno(3.5) $$  To
achieve gauge invariance of the WZW action (3.2),  in analogy with the string
case, we propose the following Yang-Mills gauge transformation rules $$
\eqalignno{      \delta B_{2n}&=- I_{2n}^1(\a, \lambda)  &(3.6) \cr
\delta C_{2n} &= I_{2n}^1(\a, \lambda)-I_{2n}^1(\b, \lambda) & (3.7) \cr} $$
Thus the problem of finding a gauge invariant coupling of the Yang-Mills field
to a $(2n-1)$-brane has been essentially reduced to finding $C_{2n}(\a,\b)$
which transforms as in (3.7).   It can be constructed as follows.

We first observe that since the Lagrangian ${\cal L}=  {\cal B}_{2n} $ is gauge
invariant up to a total derivative,  its exterior derivative is gauge
invariant, i.e. $\delta (d{\cal L})=0$.  Hence $d{\cal L}$  can be written as a
sum of separately gauge invariant pieces as follows $$       {\cal H}_{2n+1}
\equiv  d {\cal B}_{2n}  =H_{2n+1}+R_{2n+1}  \eqno(3.8) $$ where we use (3.4)
and  $$ \eqalignno{            H_{2n+1} &=  dB_{2n}+I_{2n+1}^0(\a)  &(3.9)\cr
           R_{2n+1} &=-I_{2n+1}^0(\a)  +I_{2n+1}^0(\b) +d C_{2n}(\a,\b) &
(3.10) \cr} $$   We can derive explicit formulae for expressions
$R_{2n+1}(\a,\b) $ and $ C_{2n} (\a,\b)$ which satisfy this equation in the
following way.  First introduce the following quantities. $$ \eqalign{
\a_t &= t\a+(1-t) \b   \cr            F_t &= d\a_t +{\a_t}^2  \cr
    &= t F+t(t-1)(\a-\b)^2   \cr} \eqno(3.11) $$ We then define the following
operators $$ \eqalign{               d_t &= dt {d\over dt}   \cr
l_t &=dt (\a-\b){\p\over \p F_t}  \cr} \eqno(3.12) $$ which, as shown in ref.
[8], obey the following equation $$      d_t N = (l_t d  -d  l_t ) N  \eqno
(3.13) $$ for any local polynomial $N$ in  the forms $\a_t$ and $F_t$ and the
operators $d, \ d_t$ and $l_t$.   The operator  $l_t$ is a derivation which
reduces the form degree of $N$ in $\xi$ by one and increases the form degree of
$N$ in $t$ by one, by replacing a factor of $F_t$ with $dt(\a-\b)$.  We now
choose $$     N= I_{2n+1}^0 (F_t, \a_t)   \eqno(3.14) $$ Substituting this into
(3.13), and integrating from $t=0$ to $t=1$ we obtain $$
I_{2n+1}^0(\a)-I_{2n+1}^0(\b)= \int_0^1 l_t I_{2n+2}(F_t) -d\int_0^1 l_t
I_{2n+1}^0 (F_t, \a_t) \eqno(3.15) $$ The first term on the right hand side is
manifestly gauge invariant. Thus, comparing with (3.10) we read off the
expressions $$ \eqalignno{    R_{2n+1} (\a,\b) &=\int_0^1dt  J {\partial\over
\partial F_t}  I_{2n+2}(F_t)   & (3.16)\cr  C_{2n} (\a,\b)  &= \int_0^1dt\   J
{\partial\over \partial F_t}    I_{2n+1}^0(F_t, \a_t)  &(3.17)\cr} $$ Since
$R_{2n+1}(\a,\b)$ is gauge invariant, from (3.10) we now see that the variation
of $C_{2n} (\a,\b)$ is indeed given by (3.7).

All invariants $I_{2k}(F)$  can be expressed as the products of the primitive
invariants of lower rank $ P_{2n+2}$ given by $$ P_{2n+2}(F) = {\rm tr }
F^{n+1}= d \omega_{2n+1}^0  \eqno(3.18) $$ A general formula for the
Chern-Simons form  $\omega_{2n+1}^0 $ is well known, $$ \eqalignno{
\omega_{2n+1}^0  &= (n+1)\int_0^1dt\ tr\bigg(\a F_t^n \bigg),  &(3.19) \cr} $$
where here $F_t=t F+t(t-1) \a^2$.  Some examples are  $$      \eqalignno{
    \omega_3^0 &= {\rm tr} \bigg(F\a -{1\over 3}\a^3\bigg)     &(3.20)\cr
    \omega_5^0 &= {\rm tr} \bigg(F^2\a -\h F \a^3+{1\over 10}\a^5\bigg)  &
(3.21)\cr           \omega_7^0 &= {\rm tr } \bigg(F^3\a -{2\over 5}F^2\a^3
-{1\over 5}F\a F\a^2 +{1\over 5}F\a^5  -{1\over 35}\a^7\bigg)&(3.22)\cr} $$

Using the formulae given above, we shall now work out explicitly the
expressions for $C_4$ and $C_6$,  occuring in the action for 3-branes, and
5-branes, respectively.  In the case of 3-branes,  as a starting point we
consider $$     I_6 = c_1 {\rm tr }F^3  \eqno(3.23) $$ From (3.17 ) we then
obtain the result $$    C_4 (\a,\b)= \h c_1 {\rm tr}\bigg\{ ( F\a+\a F-\a^3)\b
+ {1\over 2} \a\b\a\b - \a\b^3 \bigg\}  \eqno (3.24) $$  We can  rewrite this
result in many different ways by partially integrating and discarding total
derivatives, which drop out in  the action.  In summary,  the gauge invariant
WZW action for the 3-brane is  $$  S_4^W =\int  {\cal B}_4 ,  \eqno(3.25)  $$
where ${\cal B}_4$ is defined in (3.3).  The case of 5-branes is somewhat more
complicated. We can now consider the invariant  $$  I_8(F)= c_1 tr F^4 +c_2 (tr
F^2)^2,  \eqno(3.26)  $$ where $c_1$ and $c_2$ are arbitrary constants.  It is
easily seen that $$ I_7^0 = c_1 \omega_7^0 +c_2 (tr F^2 ) \omega_3^0
\eqno(3.27) $$ Substituting this into (3.17),   after a tedious but
straightforward calculation we find the result $$ \eqalign{    C_6(\a,\b )=&
\bigg( c_1 d^{efgh}+c_2d^{(ef} d^{gh)}\bigg)\a^g\b^h\Bigg\{ F^e F^f   +{1\over
10}f^e_{{}ab}F^f\bigg (3\b^a\b^b-4\b^a J^b+4 J^a J^b\bigg)\cr    &+{1\over
60}f^e_{{}ab}f^f_{{}cd}\bigg(3\b^a\b^b\b^c\b^d+6 \b^a\b^b\b^c J^d+ 5\b^a\b^b
J^c J^d +4\b^a J^b\b^c J^d \cr  &+6\b^a J^b J^c J^d+3  J^a J^b J^c J^d
\bigg)\Bigg\},  \cr}\eqno(3.28) $$ where $F^a = {\rm tr} (T^a F)$ and $d^{abcd}
= {\rm tr} [T^{(a} T^b T^c T^{d)}] $. The gauge invariant WZW action for the
5-brane can then be written as $$     S_6^W=\int  {\cal B}_6, \eqno(3.29)  $$
where ${\cal B}_6$ is defined in (3.2).

 Let us now turn to the case of even $p$-branes with $p=2n$.   The kinetic
action  is given in (3.1).  A rigidly $G$-invariant WZW term  requires the
existence of a rank $2n+2$ totally antisymmetric group invariant tensor, but
for   semi-simple groups these are absent until $p=4$ and for simple groups
$G$, they are absent until p = 6.  For example for $p=4$, and a group of the
form $G = G_1 + G_2$, we could take $d b_5 +\omega_3(K_1) \omega_3(K_2) =0 $ ;
for p= 6 , with $G= SU(N), N \geq 3$, we could take $d b_7 + \omega_3  \omega_5
=0 $; and for p= 8 , with $G= SO(2N), N \geq 3$, we could take $d b_9  +
\omega_3  \omega_7 =0 $.

   However,  most of the  ingredients that went into the above construction  of
a {\it locally}  $G$-invariant WZW action are only applicable for odd
$p$-branes.  For example, the nontrivially gauge invariant field strength
$H_{2n+1}$ which involves the Chern-Simons form $I_{2n+1}^0(F,\a)$ has no
analog for even $p$.  This suggests that the field $B_{2n+1}$ is inert under
Yang-Mills gauge transformations. Therefore, the methods we used for  odd
$p$-branes have to be modified. To this end, we first observe that for even
$p$-branes $b_{2n+1}$ also satisfies $db_{2n+1}+I_{2n+2}(K)=0.$ In this case,
$I_{2n+2}(K)$ can always be written as a product of an even number of primitive
Chern-Simons forms $\omega_{2k+1}(K)$.   Such factorizations  follow from the
cohomology of Lie algebras, and they can be deduced from ref.~[7].
Consequently it is always true that $b_{2n+1}$  factorizes as $$ b_{2n+1} =
b_{2i_1}\prod_{k=2}^{2q}d b_{2i_k},     \quad\quad   \sum_1^{2q}i_k=n+1-q
\eqno(3.30) $$ This suggests that we introduce $X$-dependent lower rank
antisymmetric tensor fields  $B_{2i_k}$ corresponding to each $y$-dependent one
$ b_{2i_k}$.  This furthermore suggests that we use the forms $ {\cal B}_{2n}$
as building blocks for a gauge invariant Lagrangian, since they have nice
transformation properties and contain both $B_{2 i_k}$ and $b_{2 i_k}$. We
propose the following action for $p=2n$ $$ S^W_{2n+1}= \int \Big\{ B_{2n+1}(X)+
 {\cal B}_{2i_1}\prod_{k=2}^{2q} d {\cal B}_{2i_k}\Big\},     \quad\quad
\sum_1^{2q}i_k=n+1-q   \eqno(3.31) $$ This action contains the rigid term
(3.30), and it is indeed manifestly gauge invariant, since $ {\cal B}_{2i_k}$
transforms into a total derivative.

We note that the factorization of the invariant tensor occuring on the right
hand side of (3.31) as discussed above,  can occur in some cases for odd
p-branes as well, depending on the gauge group. In such cases,  lower rank
antisymmetric tensor fields $B_{2i_k}(X)$ can again be introduced, and gauge
invariant actions of the type (3.31) can be written down.

Another generalization of the above construction is to introduce as a  factor
in the Lagrangian density the gauge invariant polynomials $I_{2i+2}(F)$ and
lower rank tensors $B_{2i +1}$ of odd degree that are taken to be inert under
the Yang-Mills transformations.  Putting all these together we arrive at a
rather general form of the locally gauge invariant WZW term which can be
written for both odd and even $p$-branes as follows $$  S^W_{2n+\epsilon}= \int
\Big\{ \sum_{ i  } c_i \epsilon B_{2i+\epsilon }(X) I_{2n - 2i}(F)+
\sum_{\{i_k\}} c_{\{i_k\}} {\cal B}_{2i_1} \ I_{2i_2} (F)\  {\cal H}_{2i_3+1}\
{\cal H}_{2i_4+1}\cdots {\cal H}_{2i_k+1}\Big\}, \quad\quad,  \eqno(3.32)  $$
where $\epsilon=0,1$  corresponding  to even and odd branes, respectively, $c_{
i } $ and $c_{\{i_k\}} $ are a  set of arbitrary constants. Here  $\{i_k\}$ is
any partition  and $q$ is any integer such that
$\sum_{k=1}^{2q+\epsilon}i_k=n+1-q$.  Without loss of generality, we can define
 ${\cal B}_{2n},\ I_{2n}(A)$ and ${\cal H}_{2n+1}$ in terms of the primitive
Chern-Simons form  $\omega_{2k+1}(F, A)$ instead of $I_{2k+1}(F, A)$.  This can
be accomplished  by field dependent redefinitions of higher rank forms
$B_{2m}(X)$ in terms of the lower rank ones. For example, in the case of
five-branes if we have the lower rank 2-form $B_2$ in addition to $B_ 6$, then
the relevant redefinition is of the form $B_6\rightarrow B_6-c_2 I_4(F)B_2$.

 \noindent{\bf 4. Comments} \bigskip In this paper we focused on generalizing
the group manifold approach to Yang-Mills couplings, with semi-simple groups,
and applying it to bosonic $p$-branes. There is clearly much scope for further
work: including $U(1)$ groups, gauging both $G_L$ and $G_R$, considering $G/H$
coset spaces instead of group manifolds, including gravitational Chern-Simons
corrections, and including supersymmetry. We do not anticipate any severe
problems in these directions. Much more problematical, in our estimation, will
be to preserve the $\kappa$-symmetry of the super $p$-branes when the
Yang-Mills couplings are included.  (For the case of string this has been done
[9]).  The solution to this latter problem is, of course, a prerequisite for
constructing the action for the heterotic 5-brane, and testing the ideas that
it might provide a dual description of the heterotic string [2,3,10].  We are
encouraged, however, by the observation that the 5-brane Chern-Simons terms
(3.27):  $$ d H_7 = d I_7 =c_1 {\rm tr} F^4 +c_2({\rm tr} F^2)^2  \eqno(4.1) $$
obtained in this paper are entirely consistent with an earlier conjecture based
on string/fivebrane duality [10]. Recall that the string one-loop Green-Schwarz
anomaly cancellation mechanism requires a correction term $B\wedge {\rm tr}
(F\wedge F\wedge F\wedge F)$ in the $D=10$ Lagrangian [11]. (For concreteness
we focus on $SO(32))$. This corresponds to a string one-loop correction to the
$H_3$ field equation, namely $$   d^*\big(e^{-\phi} H_3\big)={2\kappa^2\over 3
\alpha'(2\pi)^5} {\rm tr} F^4,   \eqno(4.2) $$ where $\phi$ is the dilaton and
$\alpha'={1\over 2\pi T_2}$ and $T_2$ is  the string tension.  But by
string/fivebrane duality $H_3$ is related to $H_7$ of the fivebrane by
$H_7=e^{-\phi} {^*}H_3 $.  Moreover, the string tension $T_2$ and the fivebrane
tension $T_6$ are quantized according to $\kappa^2T_2T_6=n\pi,\ n =$ integer.
We may thus re-interpret (4.2) as a fivebrane tree-level correction to the
$H_7$ Bianchi identity, namely [10] $$ dH_7 = n {\beta' \over 3}{\rm tr } F^4,
\eqno(4.3) $$ where $\beta'={1\over [(2\pi)^3T_6]}$.  This is consistent with
(4.1). (In the case of the string the coefficient $c_1$ in $dH_3 = c_1 {\rm tr}
F^2$ is quantized and fixed to be $c_1 = 2m \alpha', m =$ integer, by conformal
invariance.  This is also demanded by $\kappa$ symmetry.  We expect that
$\kappa$ invariance will lead to analogous restrictions on $c_1$ and $c_2$ in
(4.1).  In any case, it would appear that string/fivebrane duality requires
$c_1= n \beta' / 3$ and $c_2=0$). That the classical fivebrane considerations
of this paper should gel with quantum string effects represents, in our
opinion, further circumstantial evidence in favour of string/fivebrane duality.
\vfill\eject \bigskip \centerline{\bf REFERENCES} \bigskip \item{[1]} D.J.
Gross, J.A. Harvey, E. Martinec and R. Rohm, Nucl. Phys. {\bf B256} (1985) 253.
\item{[2]} M.J. Duff, Class. Quantum Grav. {\bf 5} (1988) 189. \item{[3]} A.
Strominger, Nucl. Phys. {\bf B343} (1990) 167. \item{[4]} E. Bergshoeff, E.
Sezgin and P.K. Townsend, Phys. Lett. {\bf 189B} (1987) 75. \item{[5]} E.
Witten, Commun. Math. Phys. {\bf 92} (1984) 455. \item{[6]} M.J. Duff, B.E.W.
Nilsson and C.N. Pope,   Phys. Lett. {\bf 163B} (1985) 343; M.J. Duff, B.E.W.
Nilsson, C.N. Pope and N. P. Warner, Phys. Lett. {\bf 171B} (1986) 170.
\item{[7]} {\it Encyclopedic Dictionary of Mathematics}, Eds. S. Iyanaga and Y.
Kawada (MIT Press, 1980). \item{[8]} J Manes, R. Stora and B. Zumino, Commun.
Math. Phys. {\bf 102} (1985) 157. \item{[9]}  R. Kallosh, Physica Scripta {\bf
T15}(1987) 118; Phys. Lett. {\bf  176B}(1986) 50.  See also, E. Bergshoeff, F.
Delduc and E. Sokatchev,  Phys. Lett. {\bf 262B} (1991) 444. \item{[10]} M.J.
Duff and J.X. Lu, Phys. Rev. Lett {\bf 66} (1991) 1402; Nucl. Phys. {\bf B357}
(1991) 534. \item{[11]} M.B. Green and J.H. Schwarz, Phys. Lett. {\bf 173B}
(1984) 52. \vfill\eject \end